\documentclass{ieeeaccess}

\usepackage{listings}
\usepackage{cite}
\usepackage{amsmath,amssymb,amsfonts}
\usepackage{algorithmic}
\usepackage{graphicx}
\usepackage{textcomp}
\usepackage{xcolor}
\usepackage{listings}
\usepackage{float}
\usepackage{flafter} 
\usepackage{caption}

\definecolor{lightgray}{gray}{0.95}
\definecolor{blue}{rgb}{0,0.2,1} 
\definecolor{red}{rgb}{1,0,0} 
\definecolor{orange}{rgb}{1,0.5,0} 
\definecolor{cyan}{rgb}{0,0.8, 0.8} 
\definecolor{mustard}{rgb}{0.85, 0.65, 0.13}
\definecolor{green}{rgb}{0,0.6,0} 
\definecolor{purple}{rgb}{0.7,0,0.7} 
\definecolor{accessblue}{cmyk}{1,0.45,0,0.18}

\def\BibTeX{{\rm B\kern-.05em{\sc i\kern-.025em b}\kern-.08em
    T\kern-.1667em\lower.7ex\hbox{E}\kern-.125emX}}

\begin{document}
\history{Date of publication xxxx 00, 0000, date of current version xxxx 00, 0000.}
\doi{10.1109/TQE.2020.DOI}
\title{Benchmarking the ability of a controller to execute quantum error corrected non-Clifford circuits}
\author{\uppercase{Yaniv Kurman\authorrefmark{1}, Lior Ella\authorrefmark{1}, Ramon Szmuk\authorrefmark{1}, Oded Wertheim\authorrefmark{1}, Benedikt Dorschner\authorrefmark{2}, Sam Stanwyck\authorrefmark{2},  and Yonatan Cohen}.\authorrefmark{1}}
\address[1]{Quantum Machines Inc., Tel Aviv, Israel (email: author@boulder.nist.gov)}
\address[2]{NVIDIA Corp, Santa Clara, CA, USA}

\markboth
{Author \headeretal: Preparation of Papers for IEEE Transactions on Quantum Engineering}
{Author \headeretal: Preparation of Papers for IEEE Transactions on Quantum Engineering}

\corresp{Corresponding author: Yaniv Kurman (email: yanivk@Quantum-machines.co).}

\begin{abstract}
Reaching fault-tolerant quantum computation relies on the successful implementation of non-Clifford circuits with quantum error correction (QEC). In QEC, quantum gates and measurements encode quantum information into an error-protected Hilbert space, while classical processing decodes the measurements into logical errors. QEC non-Clifford gates pose the greatest computation challenge from the classical controller’s perspective, as they require mid-circuit decoding-dependent feed-forward —modifying the physical gate sequence based on the decoding outcome of previous measurements within the same circuit. In this work, we introduce the first benchmarks to holistically evaluate the capability of a combined controller-decoder system to run non-Clifford QEC circuits.  We show that executing an error-corrected non-Clifford circuit, comprised of numerous non-Clifford gates, strictly hinges upon the classical controller-decoder system. Particularly, its ability to perform decoding-based feed-forward with low-latency, defined as the time between the last measurement required for decoding and the dependent mid-circuit quantum operation. We analyze how the system’s latency dictates the circuit’s operational regime: latency divergence, classical-controller-limited runtime, or quantum-operation-limited runtime. Based on this understanding, we introduce latency-based benchmarks to set a standard for developing QEC control systems as an essential components of fault-tolerant quantum computation.
\end{abstract}


\maketitle
\section{Introduction}
\label{sec:introduction}
\PARstart{Q}{uantum} error correction (QEC) \cite{shor1996fault, knill1998resilient} stands as the clearest path for achieving the advantage of quantum computing, enabling the resolution of significant problems such as simulating complex quantum systems \cite{feynman2018simulating}, factorization \cite{shor1999polynomial}, and more. In expected QEC implementations, quantum logic is encoded across several physically separated qubits, with repeated local parity measurements (stabilizer rounds) facilitating the detection and correction of local physical errors \cite{gottesman1997stabilizer}. Although increasing the number of physical qubits introduce additional errors, this expansion can exponentially reduce logical quantum errors if the physical errors are below a certain QEC code-dependent threshold \cite{preskill1998reliable,campbell2017roads}, opening the path toward useful quantum computation. 

The successful execution of QEC codes depends on the performance of several key components. First is the performance of the qubits themselves, whose physical error rates should be sufficiently smaller than the QEC error threshold. Secondly, the quantum controller which executes the quantum computation should perform all quantum gate operations with optimal fidelities. Thirdly, a successful QEC implementation requires a quantum error decoding unit (decoder). The decoder maps the physical quantum measurements into quantum errors using a classical algorithmic procedure \cite{demarti2024decoding}. Since QEC decoding algorithms are significantly more complex than the current embedded classical processing in a quantum control unit \cite{ofek2016extending,devulapalli2024quantum, baumer2024efficient, foss2023experimental, ella2023quantum}, state-of-the-art QEC experiments still make use of independently developed quantum control and decoding units \cite{ai2024quantum, bluvstein2024logical}. 

High performance and tight integration of all classical components are essential for achieving any quantum advantage with QEC circuits, that is, for executing non-Clifford circuits. Non-Clifford gates play a crucial role in quantum computation, as they are necessary to complete a universal gate set when combined with Clifford operations \cite{gottesman1998heisenberg, dawson2005solovay}. While Clifford gates alone can be efficiently simulated classically due to the Gottesman-Knill theorem \cite{gottesman1998heisenberg}, non-Clifford gates introduce the computational complexity required for quantum advantage. However, implementing these gates in a fault-tolerant manner requires feed-forward operations, where mid-circuit quantum gates depend on the real-time decoding of previous measurement outcomes \cite{riesebos2017pauli}. 

The necessity for low-latency decoding-dependent feed-forward imposes stringent latency and throughout constraints on the decoding system to prevent catastrophic backlogs within non-Clifford circuits \cite{terhal2015quantum} , and low-latency controller-decoder integration. Recent experimental efforts have demonstrated reaching these constraints in real-time decoding\cite{ai2024quantum}, successfully integrating controller-decoder feed-forward \cite{bluvstein2024logical, caune2024demonstrating}, and ultra-fast decoding hardware with sublinear complexity and microsecond-scale latency \cite{liyanage2023scalable}. Complementary to these, dedicated hardware architectures have been developed \cite{das2022afs}, and classical computational bottlenecks have been analyzed \cite{maurya2024managing}. These works underscore the developments within each domain toward building complete classical systems capable of running fault-tolerant quantum computation.

In this manuscript, we introduce the first holistic benchmark designed to evaluate the performance of the entire classical stack in a quantum computer, including the controller, decoder, and their integration. To emphasize its holistic nature, we denote this part of the computer as a “controller-decoder unit” (CDU). 

Our holistic approach serves as a higher-level benchmark compared to previous evaluations that address individual components separately. To date, quantum control units and QEC decoders have been evaluated using distinct criteria. Evaluation of quantum control units is centered around the analog output performance and parallel quantum-classical processing \cite{ella2023quantum, wack2110quality}. QEC decoders are primarily evaluated based on their ability to achieve decoding times per stabilizer round shorter than the round duration itself \cite{battistel2023real}, preventing catastrophic backlogs and diverging decoding durations \cite{terhal2015quantum}. The primary goal of decoding hardware development has been to minimize decoding time \cite{wu2023fusion, liyanage2023scalable, barber2025real}. However, these separate evaluations fail to establish a unified criterion for assessing whether an integrated controller-decoder system can successfully execute QEC algorithms.

The benchmarks which we introduce here focus on the ability of the CDU to apply decoding-based, low-latency, feed-forward operations during circuit execution, where the feed-forward latency (FFL) is defined by the time interval between the pulse of a feed-forward gate (mid-circuit gate modification) and the last measurement it relies on. Our analysis shows how the evolution of the FFL during the QEC computation determines whether the CDU will reach a continuously increasing (diverging) latency value, thus rendering the quantum computer inoperable, or a steady-state FFL. The steady-state FFL is defined as an equilibrium point between the time it takes to decode a certain dataset and the time in which this to-be-decoded dataset is created. This latency depends on the decoding time, the controller-decoder communication channel, and the controller’s operation speed. Moreover, its value is a crucial performance metric since it determines the duration of a non-Clifford gate and thus its fidelity. Our suggested benchmarks will verify the CDU’s capability to apply decoding-dependent quantum gates while parallelizing the decoding itself with quantum processing and evaluate the overall steady-state duration, under near-term settings with only one or two logical qubits. 

Our manuscript is structured as follows. Section II serves as an introduction to QEC-based quantum computation with the well-known surface code. Section III presents the CDU as a control unit, which functions as the cornerstone for the execution of QEC-based quantum computations. Section IV elucidates the role of the FFL in QEC computation and shows how to determine its steady-state behavior using a dynamical system approach. In section V, we suggest two benchmarks, exemplified by simple logical operations, and evaluated given different system parameters. Finally, in Section VI, we discuss the long-term scalability requirements of the CDU as quantum hardware scales, and propose additional possible benchmark.

\section{QEC QUANTUM COMPUTATION WITH SURFACE CODES}

The surface code \cite{dennis2002topological, fowler2012surface} stands out among the different QEC codes because of its high error threshold of \(~1\%\), orders of magnitude higher than other codes, and simple physical requirements. Many aspects of the surface code have been developed in detail, including fault-tolerant computation techniques \cite{horsman2012surface,chamberland2022universal,litinski2019game,higgott2023improved}, high-fidelity magic state preparation techniques \cite{li2015magic,gidney2023cleaner}, decoding algorithms \cite{kolmogorov2009blossom,delfosse2021almost,chubb2021general,das2022lilliput, meinerz2022scalable,wu2023fusion,higgott2022pymatching,battistel2023real}, simulation tools \cite{gidney2021stim}, distillation schemes \cite{litinski2019game,haah2018codes,litinski2019magic}, and scaling proposals \cite{chamberland2022universal,litinski2019game}. This motivates us to define the CDU benchmarks with the surface codes as a representative example. Nevertheless, we keep in mind that the benchmarks below can be easily modified to any stabilizer code.

We start with a brief introduction to QEC quantum computation with the surface-code. The surface code implements each logical qubit with a set of parity checks that can be tiled onto a square lattice. In the qubit-efficient rotated surface implementation, the quantum information of a single qubit (logical qubit) is encoded in a single surface with \(d^2\) physical qubits called data qubits (black circles in Fig. 1a), where \(d\) is the distance of the QEC code. In addition to the data qubits, the logical qubit requires \(d^2-1\) ancillary physical qubits which perform Pauli \(Z\) or Pauli \(X\) (blue and purple circles in Fig. 1a, respectively) local parity measurements of their nearest-neighbor data qubits. These Pauli \(Z\) and Pauli \(X\) measurements track the presence of local bit or phase flip errors, respectively, on the data qubits. 

The \(+1\) eigenstate of all measured Pauli operators defines the Hilbert subspace of the logical qubit (called the code space) \cite{gottesman1997stabilizer}, which the data qubits collapse into up to a local bit or phase flip. This collapse to the code space is called stabilization, the Pauli operators that are probed are called stabilizers, and the act of probing is called a stabilizer round which is constantly repeated during all fault-tolerant logical quantum gates of the surface codes. For example, Fig. 1b presents the required physical gate sequence that implements a multi-surface parity check (i.e., lattice-surgery) fault-tolerantly by repeating the QEC stabilizer round \(d\) times.

\begin{figure}[t]
    \centering
    \includegraphics[width=\linewidth]{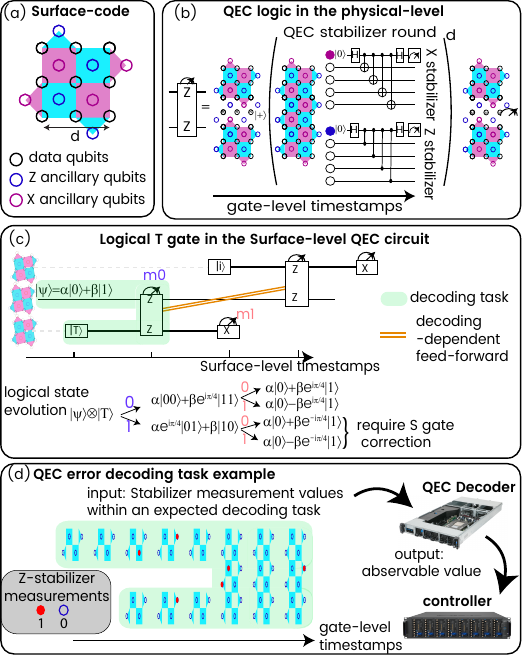}
    \caption{\textbf{Example of non-Clifford QEC computation with surface codes. (a)} A logical qubit in the surface code, implemented with \(d^2\) data qubits (black) and \(d^2-1\) ancilla qubits (blue and purple) which are used for the stabilizer measurements. \textbf{(b)} The physical-level QEC quantum circuit, which implements a logical \(ZZ\) parity measurement through initialization of data qubits which connect two surfaces, d repeated stabilizer rounds for the elongated surface, and measurement of the connecting data qubits. We note that this operation is fault-tolerant due to the d repetitions of the stabilizer round. \textbf{(c)} The surface level QEC circuit, which implements a single non-Clifford (\(T\)) gate. This is done, for example, using two ancillary surfaces and the native surface-level gates which implement fault-tolerantly measurement-based quantum computation. To implement a \(T\) gate rather than a \(T^{-1}\) (up to a Pauli \(Z\) correction), a logical feed-forward in the form of an S gate is needed, depending on the \(m_0\) result, as we show in the logical state evolution. Decoding is required to find \(m_0\) (decoding task marked in green), and thus the feed-forward (orange) which implements the \(S\) gate (using a \(\lvert i\rangle=\lvert 0\rangle+i \lvert 1\rangle\) state) is decoding-dependent. \textbf{(d)} An example of the decoding task in (c) for a \(d=3\) code. The stabilizer measurement results (\(Z\)-stabilizers) which correspond to the logical measurement outcome of interest (logical \(ZZ\) parity measurements) are sent to a QEC decoder which determines and sends to the controller its best estimation to the logical measurement value.}
    \label{fig1}
\end{figure}

An established implementation of quantum information processing with QEC uses measurement-based quantum computation. The fault-tolerant logic gate-set includes single-logical-qubit (i.e., single-surface) measurements and initializations in the \(X\) or \(Z\) basis, as well as multi-logical-qubit parity measurements in the \(Z/X\) basis (e.g., \(ZZ, XX, ZX\) for two surfaces) \cite{horsman2012surface,chamberland2022universal,litinski2019game,higgott2023improved}. The commonality between all the above is that their physical implementation can be verified with (or implemented by) the QEC code’s stabilizers, where a padding of \(d\) stabilizer rounds before, after, or during the logical gate are needed for a fault-tolerant implementation.

To complete to a universal gate set and reach quantum advantage (Gottesman-Knill theorem) \cite{gottesman1998heisenberg}, it is required to perform surface initialization in a logical magic-state, namely a state which is not a logical Pauli eigenstate. Such preparation is non-fault-tolerant and will eventually require magic state distillation \cite{litinski2019game,haah2018codes,litinski2019magic}. Once this magic state is prepared, with or without distillation, it is used to implement a non-Clifford gate on a logical qubit.  

Fig. 1c illustrates a QEC computation example that implements a non-Clifford \(T=\text{diag}(1,e^{i\pi/4})\) gate on a general logical qubit \(|\psi\rangle=\alpha|0\rangle+\beta|1\rangle\) using an ancillary surface initialized in the logical \(|T\rangle=|0\rangle+e^{i\pi/4}|1\rangle\) magic-state. 

Each measurement outcome during the measurement-based computation has a \(50\%\) chance of being \(0\) or \(1\), which dictates a required logical-circuit modification. A correction within the Pauli group (\(X, Y,\) or \(Z\)) in the logical circuit can propagate in software without modifying the logical circuit (thus avoiding additional errors and delays), and therefore does not require logical circuit modifications \cite{fowler2012surface}. However, a correction in the form of a logical \(S=\text{diag}(1,i)\) gate cannot propagate through a non-commuting non-Clifford gate (as explained in SM Section S1). Consequently, feed-forward must be applied to implement a logical non-Clifford gate (in our example, if \(m_0\) is \(1\)), and moreover, it is decoding-dependent.

\begin{figure*}
    \centering
    \includegraphics[width=\linewidth]{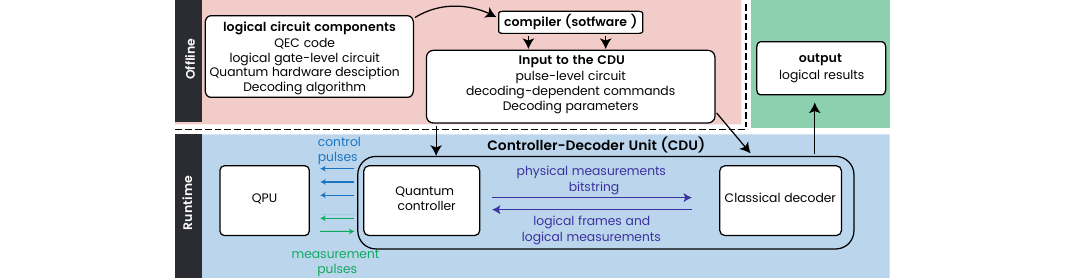}
    \caption{The CDU’s central role in QEC quantum computation. The CDU is a runtime hardware component which includes both the classical decoder and the quantum controller that directly interacts with quantum processing unit (QPU). Here we show an implementation for running a QEC logical circuits: an offline software compiler converts the QEC code, logical gate-level circuit, quantum hardware description, and decoding algorithm into the input of the CDU: a pulse-level circuit and decoding parameters. During runtime, the physical circuit is executed by the CDU with syndrome flow from the controller to the decoder while the decoding results, in the form of logical measurement results and logical frames, are both returned to the quantum controller and sent to the user.}
    \label{fig2}
\end{figure*}

The effectiveness of QEC in reducing logical errors depends on accurate error decoding. Decoding is a classical algorithmic process that analyzes all physical measurements acquired up to a certain point and converts this data into logical measurement outcomes and logical Pauli corrections for the corresponding logical qubits. Decoding during the evolution of the circuit is essential because some Pauli errors cannot propagate through non-Clifford gates without transforming into non-Clifford errors (see SI Section S1). Therefore, it is crucial to track physical errors within logical qubits online to implement correct quantum logic. 
The field of decoding algorithms is under intense research \cite{battistel2023real} with available algorithms exhibiting trade-offs between high accuracy \cite{chubb2021general}, speed \cite{das2022lilliput}, scalability \cite{meinerz2022scalable}, and potential for implementation in available hardware \cite{liyanage2023scalable}. 

Fig. 1d demonstrates a decoding procedure. The decoder's input includes all physical measurements that can track an error in a logical measurement outcome (in this case, an error in the \(ZZ\) multi-qubit measurement from Fig. 1c), and the output is the logical measurement value, possibly accompanied by additional metadata such as logical qubit Pauli frame flips (\(I, X, Y,\) or \(Z\)). The decoder's output is then sent to the quantum control unit, which executes the physical implementation of the decoding-dependent logical gates.

The benchmarks below assess the time from the physical execution of a logical measurement until the time the controller issues a conditional pulse based on the decoded logical measurement outcome. Closing this loop with low latencies motivates the transition from viewing the decoder and controller separately to adopting a holistic perspective of a CDU.
\section{The CDU}

We define the controller-decoder unit (CDU) as a classical hardware element designed to enable full execution of QEC quantum computation. The CDU must conform to several requirements: (i) executing QEC quantum gate sequences via analog signals to the quantum processors, (ii) performing error decoding, and (iii) integrating the controller and decoder to execute quantum gates based on decoding outcomes. Importantly, we do not limit the CDU to any specific quantum hardware, classical hardware implementation, QEC code, decoding algorithm, micro-architecture, or controller-decoder communication protocol.

Fig. 2 illustrates the central role of the CDU in QEC quantum computation. Before execution (offline), it is essential to define the CDU tasks, which include the QEC code, the target quantum logical gate-level circuit, the decoding algorithm, and the physical description of the quantum computer. This information is compiled into a set of inputs for the CDU. The quantum controller receives the pulse-level circuit to run on the quantum processing unit (QPU), including optimized pulse sequences, classical decision-making, and decoding-dependent conditional gates. The CDU's decoder receives the expected data transfer, parameters such as the matching graph dimensions, which can be static or dynamic. 

Once all components of the CDU are ready, circuit execution begins (runtime). Physical quantum measurement results (ancillary and data-qubit measurements) are sent to the decoder, which returns logical frames and logical measurement results to the controller and the user. The benchmarks below aim to verify the CDU ability to perform these tasks.

Current state-of-th-art quantum control units can already fulfill their requirements due to their flexibility in executing quantum logic with mid-circuit control flow capabilities such as conditional commands, loops, and other branching at nanosecond timescales, enabling agile and non-deterministic quantum circuits \cite{ella2023quantum}.

Other CDU components are also well-developed, though tailored for specific tasks. Several decoding algorithms such as Union-find  \cite{delfosse2021almost}, minimum-wight-perfect-matching (MWPM) \cite{kolmogorov2009blossom}, and neural-network decoders \cite{meinerz2022scalable} were implemented in FPGA \cite{liyanage2023scalable} or dedicated hardware \cite{das2022afs,barber2025real}, achieving an average decoding latency per round of below 1 microsecond \cite{caune2024demonstrating}. In addition, small decoding tasks benefit from a look-up table decoder \cite{das2022lilliput} which can be implemented directly in the controller, while large decoding tasks might leverage a parallelizable algorithm such as fusion blossom \cite{wu2023fusion}, which is suitable for GPU implementation. 

Finally, an essential aspect of the CDU that affects its performance is its microarchitecture (its detailed and dedicated hardware structure). Recent studies have shown that a well-designed decoder microarchitecture can overcome assumed decoding tradeoffs, reaching fast, accurate and scalable performance \cite{das2022afs}. Novel controller microarchitectures have been proposed to optimize and enable scalable QEC control \cite{fu2019control}. Altogether, the diversity of each component highlights the need for a holistic CDU benchmark.

\section{THE ROLE OF FEED-FORWARD LATENCY IN QEC QUANTUM COMPUTATION}

In Section II, we discussed the necessity of conditional decoding-dependent feed-forward for implementing a T gate with surface codes. This requirement extends to the implementation of any non-Clifford gate in any QEC computation that is measurement-based, for example in \cite{litinski2019game}. A critical question arises: what are the requirements for feedforward latency (FFL), and what is the impact of excessive latency? This chapter provides a general analysis of FFL under various classical parameters, applicable to any QEC stabilizer code that requires decoding-dependent feed-forward for fault-tolerant quantum computation.

\subsection{THE DEPENDEINCIES BETWEEN DECODING TASKS IN NON-CLIFFORD QEC CIRCUITS}

To successfully execute a non-Clifford QEC circuit, the circuit is structured into distinct decoding tasks, where each non-Clifford gate includes at least one decoding task that modifies the gate sequence according to the decoding outcome\cite{kurman2024controller, maurya2024managing}. This modification introduces a critical dependency between a decoding task and its preceding tasks, as the to-be-decoded sequence is influenced by preceding decoding result. Consequently, the structure and size of each task depends on the outcomes of a prior decoding task. More importantly, since QEC gate sequences must continuously run to preserve the encoded quantum logic, the dependent task continues to grow until the decoding of the preceding tasks ends. This dependency is the root cause of the potential catastrophic backlog\cite{terhal2015quantum}, underscoring the necessity of latency-based benchmarks designed to minimize this accumulation. 

To illustrate this critical dependency in a specific quantum circuit, we analyze a circuit consisting of two consecutive non-commuting non-Clifford gates (Fig. 3a) using the limited quantum resources of a single ancillary surface. The surface-level execution of the circuit (Fig. 3b) employs the ancillary logical qubit (surface) to implement the non-Clifford gates and their conditional corrections fault-tolerantly. Importantly, since the non-Clifford gates do not commute, the feed-forward of the first gate (T) must be applied before executing the second non-Clifford gate (\(X^{1/4}\)). Thus, the computational logical qubit will remain idle (creating more measurements to be decoded) until the end of the decoding of the first logical parity measurement (ZZ lattice surgery). To execute the logic from Fig. 3(a), the logical parity measurement modifies the initialization of the ancillary surface, either in \(|i\rangle=|0\rangle+i|1\rangle\) for the S gate correction or in the magic state \(|X^{1/4}\rangle=|+\rangle+e^{i\pi/4}|-\rangle\) to start the next non-Clifford gate.

Fig. 3c presents the division of the logical circuit from Fig. 3b into different decoding tasks, illustrated in a two-dimensional section of the space-time view of the code, where the space axes correspond to the location of an error event (as Fig. 6 in \cite{higgott2023improved}). Each decoding task ends at the end of the logical measurement upon which the decoding-dependent feed-forward depends on. The first decoding task (task 0, yellow) starts in the beginning of the circuit and is responsible for determining the first ZZ measurement outcome. The endpoint of the decoding task is not well-defined since a set of errors might be fully detectable only after the ZZ measurement has ended \cite{higgott2023improved}. As a result, also the beginning of the second decoding task (decoding task 1, blue) should receive some boundary conditions from task 0. 

Fig. 3d presents how the size (in terms of syndromes to be decoded) of the task 1 is determined by the first decoding task. If the first feed-forward is applied, then the second decoding task will include an additional fixed amount of the syndromes of the S gate correction. Moreover, the size of task 1 is dependent on the first FFL (denoted as \(L^{(0)}\)) since the data to be decoded in task 1 continues to be acquired while idling \(|\psi\rangle\) during the decoding of the first task. This dependency between tasks highlights the importance of the FFL: a delayed \(L^{(0)}\) increases the challenge of the second decoding task, resulting in an increased \(L^{(1)}\) and so on. In the analysis below, we explain the requirements on the CDU to avoid such diverging effects.

\begin{figure}[H]
    \centering
    \includegraphics[width=\linewidth]{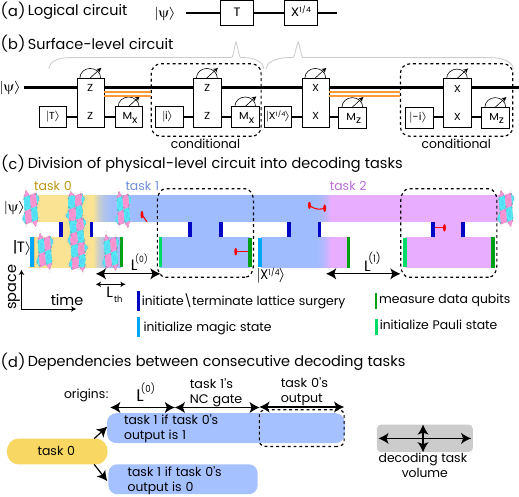}
    \caption{Compiling the decoding tasks and their dependencies from a logical circuit. (a) A logical circuit containing two non-commuting non-Clifford gates. (b) The fault-tolerant surface-level circuit implementing (a) with surface codes using a single ancillary surface. The dashed square denotes the feed-forward conditional logical gates of each non-Clifford gate. (c) The two-dimensional space-time view of the circuit in (b). Colors represent separate decoding tasks, ending each task with its logical measurement upon which a conditional feed-forward operation is applied. In this case, the decoding outcome of the lattice surgery between a magic state surface and the computation surface determines a feed-forward circuit. The FFL (\(L\)) delays the circuit if it exceeds a threshold latency \(L_{th}\), set by the number of rounds required to establish the measurement result of the ancillary surface. We note that the boundary conditions between decoding tasks are necessary to match (red lines) syndromes (red circles) between tasks. (d) Decoding task 0 determines the size of decoding task 1. The origins of the decoding task 1 volume include the latency of task 0 (\(L^{(0)}\)), a fixed volume corresponding to the non-Clifford (NC) gate of task 1 , and a conditional volume which depends on the output value of decoding task 0.}
    \label{fig3}
\end{figure}

Not all FFL values necessarily increase the size of the subsequent decoding task. In Fig. 3c, limiting to a single ancillary surface creates a threshold latency \(L_{th}\). If the controller is ready to apply the feed-forward before \(L_{th}\), the FFL will be \(L_{th}\) and it will not increase the size of the next decoding task. In this case, the classical controller waits for the ancillary surface measurement before re-initializing it. Here, \(L_{th}\) is determined by \(d\) stabilizer rounds needed for a fault-tolerant surface measurement. A similar \(L_{th}\) value appears when using two ancillary surfaces \cite{litinski2019game} (see SI section 2 and Figures S1, S2). More generally, every non-Clifford circuit and number of ancillary surfaces will determine a different \(L_{th}\) for each decoding task \cite{kurman2024controller} (For example, \(L_{th} =0\) in the circuit from Fig. 1c).

\begin{figure*}
    \centering
    \includegraphics[width=\linewidth]{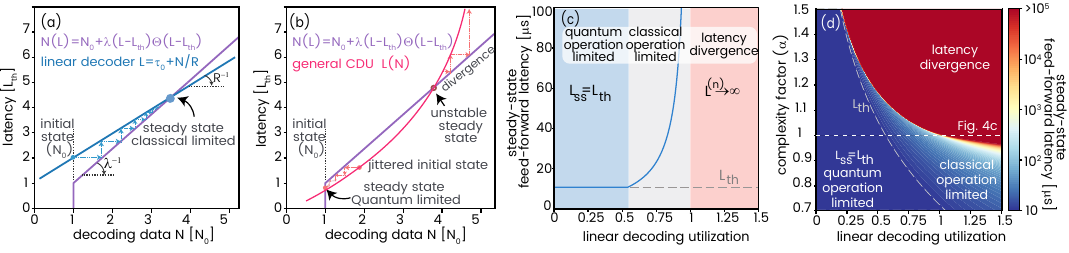}
    \caption{\textbf{Simulations of the feed-forward latency, its steady-state, and the CDU computation regimes. (a-b)} A dynamical system analysis for a linear decoder (a) and a general decoder (b), presenting the dynamics of the system using the curves of \(L(N)\) (blue and red) and \(N(L)\) (purple) from the initial state of \(L(N_0)\) until reaching a steady-state when the curves intersect. The steady-state latency can show a quantum-limited operation when converging to \(L_{th}\), classical limited operation when the steady-state is stable and above \(L_{th}\), or a latency diverging regime when the intersection does not exist or when the FFL is larger than an unstable steady-state point. \textbf{(c-d)} The steady-state FFL for a linear decoding model (c) or for different decoding parameters (d). All simulation parameters are presented in Table S1, and latency plots for various values of \(L_{th}\) are presented in Fig. S3.}
    \label{fig4}
\end{figure*}

We note that the example circuit shows a magic states prepared in a single surface-level timestamp (\(d\) stabilizer rounds), typically with a non-fault-tolerant state injection \cite{li2015magic, gidney2023cleaner}. A noisy magic state can suffice for shallow non-Clifford logical circuits \cite{kurman2024controller}, in partially fault-tolerant circuits \cite{akahoshi2024partially}, or for magic state distillation circuits \cite{litinski2019magic}. This process outputs a low-error logical initialization using many noisy magic state initializations (at least 15 for the 15-to-1 distillation process) and in a qubit-limited case (e.g., in  \cite{litinski2019game}) includes also non-Clifford gates which require feed-forward. Therefore, the magic-state distillation circuit includes a similar connection and dependency between decoding tasks as in Fig. 3, and our analysis and benchmarks below will also assess the CDU’s capability to run a magic-state distillation procedure.

\subsection{ANALYSIS OF THE CDU SYSTEM CAPABILITY TO EXECUTE NON-CLIFFORD CIRCUITS}

We established that non-Clifford quantum computation with QEC consists of dependent decoding tasks, where one task's FFL determines the next task's size. This section explains the CDU's required properties for running non-Clifford circuits on a given quantum system, showing how these properties lead to qualitatively different operation regimes that determine the feasibility of logical quantum computation. 

To elucidate the origins of the various regimes, we conduct a dynamical system analysis \cite{alligood1998chaos}. The CDU behavior is considered dynamical since the FFL of decoding task \(n\), \(L^{(n)}\), is connected to \(L^{(n-1)}\) through the number of syndromes that the decoder is required to analyze in task \(n\), \(N^{(n)}\). This connection motivates us to use the broad tool set of dynamical system analysis to derive the behavior of the CDU for a given quantum system, with variables \(N(L)\), the number of syndromes the quantum computer generates for a given FFL, and \(L(N)\), the FFL of the CDU given the number of syndromes.

Figures 4a-b show examples of using dynamical system analysis to derive the CDU’s behavior, with a linear decoder (blue line in Fig. 4a), where decoding time is linearly dependent on the number of syndromes, and a varying complexity decoder (red curve in Fig. 4b). The system’s initial state (that includes \(N_0\) syndromes) evolves (arrows) until reaching a stable steady-state point where the curves of \(L(N)\) and \(N(L)\) (purple curve in Figures 4a,b) intersect at a steady-state FFL (\(L_{ss}\)). This point determines the time required to fully execute each non-Clifford gate when the system reaches its steady-state operation point. Since the logical error depends on the size of the decoding task, \(N(L_{ss})\), the steady-state point will eventually determine the logical circuit fidelity since the logical coherence depends on the number of stabilizer rounds in the surface idling \cite{ai2024quantum}.

The \(L_{ss}\) value determines three conceptually different operation regimes. (i) Classical-operation limited regime where the system reaches a stable steady-state FFL larger than the threshold latency (\(L_{ss} > L_{th}\)). In this regime (Fig. 4a), the CDU latency limits the quantum computation logical clock but can still enable many non-Clifford gates. (ii) Quantum-operation limited regime, where the steady-state FFL equals the threshold latency (\(L_{ss} = L_{th}\)), as the stable steady-state point in Fig. 4b. In this regime, the complete circuit time depends solely on the quantum operations. (iii) Latency divergence regime, where the FFL continuously increases, making scalable quantum computation infeasible, as \(\lim_{n\to\infty} L^{(n)} \to \infty\), shown in the unstable steady-state point in Fig. 4b.

The operation regime can be derived using the following analysis. Consider a quantum system generating decoding data at a rate \(\lambda\) (syndromes per second), leading to a linear connection \(N^{(n)}(L^{(n-1)})= N_0+\lambda(L^{(n-1)}- L_{th})\Theta(L^{(n-1)}- L_{th})\), with \(N_0\) denoting the number of syndromes when the FFL is \(L_{th}\) (which for simplicity we assume independent of \(n\)), and \(\Theta()\) is a step function. Then, we can examine a linear latency model \(L(N)=\tau_0+N/R\), with throughput rate \(R\) in decoded syndromes per second, and \(\tau_0\) for the decoder-controller communication and decoder bring-up latencies. Fig. 4a shows that this linear system can reach the steady-state point only if the slope of \(L(N)\) is smaller than the slope of \((N(L))^{-1}\), i.e., if \(\lambda/R<1\). This inequality (and Figs. 4a,b) provides a mathematical (and graphical) view of the backlog preventing condition from \cite{terhal2015quantum}. 

The ratio between the syndrome generation rate from a quantum system over the decoder's throughput describes the decoder’s utilization \(U=\lambda/R\). The utilization is a well-known parameter in communication theory, where a utilization above 1 indicates an overflown channel. Fig. 4c shows \(L_{ss}\) as a function of \(U\) (calculated analytically in SI section S3) for a linear decoder. When \(L(N_0)>L_{th}\), the steady-state latency grows with the utilization and diverges as the utilization approaches 1. When \(U>1\) (red-zone), the latency diverges so that the classical behavior prevents any quantum calculation. Conversely, when \(L(N_0)<L_{th}\) (and \(U<1\)), reducing \(U\) does not change \(L_{ss}\), corresponding to an intersection when \(N(L)\) is vertical in Fig. 4a.

For a general CDU latency behavior (pink in Fig. 4b), the system will converge into a stable steady-state if \(U(N)=\lambda/ R(N)<1\) for all \(N\), with the throughput rate dependent on \(N\) via \(R(N)=(\partial L(N)/\partial N)^{-1}\). If \(U>1\) at the intersection point, the steady state is unstable and will cause latency divergence once a single FFL is higher than the unstable point. Realistic decoders show an expected FFL behavior of the form \(L(N)=\tau_0+\tau_1N^\alpha\) (e.g., \cite{wu2023fusion, barber2025real,higgott2025sparse,liyanage2023scalable}), where \(\alpha\) is the complexity factor of the decoder, and \(\tau_1\) is a pre-factor, making \(\lambda\tau_1\) the linear decoder utilization. Fig. 4d shows the steady-state FFL under variation of the complexity and the linear decoder utilization, emphasizing the importance of near-linear or sublinear complexity decoding algorithms to avoid latency divergence.

Overall, using dynamical system analysis and graphical representations enables intuitive analysis of the operational boundaries of the CDU with any quantum hardware and QEC codes. In the following section, we utilize this analysis to determine our benchmark values in different systems. These examples can extend to assess system dynamics amid significant latency or syndrome fluctuations, employing a probabilistic methodology to examine latency divergence likelihood. The dynamical-system approach is applicable to any non-linear behaviors of \(N(L)\) and \(L(N)\), correlative connections between the two, and cases where \(L_{th}\) varies between decoding tasks. In these cases, the existence and value of a stable steady-state FFL will determine the CDU's capability for useful quantum computation.

\section{CDU BENCHMARKS: STEADY-STATE FEED-FORWARD LATENCY}

After recognizing the pivotal influence of classical control and computational performance on fault-tolerant QEC computations, we propose two benchmarks aimed at evaluating the CDU’s capability to support such computations in the near and medium term. These benchmarks focus on latency benchmarking, measuring the time from the last input of measurement signals to the controller until the first decoding-dependent conditional output signal from the controller. Specifically, the duration it takes from a set of measurements that implement a logical measurement until a conditional set of feed-forward operations is performed to implement a logical feed-forward on a logical qubit, based on the error decoding of a logical measurement outcome. 

Each latency benchmark below relates to a steady-state FFL that encompasses three key aspects in a single parameter: (i) the capacity to simultaneously execute quantum operations alongside the decoding process; (ii) the capability of the CDU to avoid divergence in the decoding task and reach a steady-state point (i.e., good enough decoding throughput); and (iii) the speed of the decoder, the controller, and their integration through the steady-state FFL value.

To create well-defined and rigorous benchmarks, we focus on a specific representative configuration: a QEC rotated surface code of distance-5 and distance-11 with a stabilizer round cycle time of 1 microsecond, a union-find decoding algorithm \cite{delfosse2021almost}, physical error rates of \(0.5\%\) (two-qubit depolarization and single-qubit measurement), and a logical circuit that is uniquely defined in each of the two benchmarks below. 

We define logical circuits simple enough to keep the benchmarks implementable in the near term, using only Clifford gates and with a logical Pauli feed-forward which should occur after every decoding task. We require each physical measurement within the logical gates to be randomly generated RF (radio-frequency) pulses according to the error model so that the benchmark will include the signal processing involved in typical superconducting transmon state estimation (demodulation, integration, and threshold-based discrimination). The benchmarks can then be evaluated using the CDU only, without a connection to quantum hardware.

\subsection{NEAR-TERM BENCHMARK: STEADY-STATE INTER-CIRCUIT FEED-FORWARD LATENCY (SIFL)}

\begin{figure*}
    \centering
    \includegraphics[width=\linewidth]{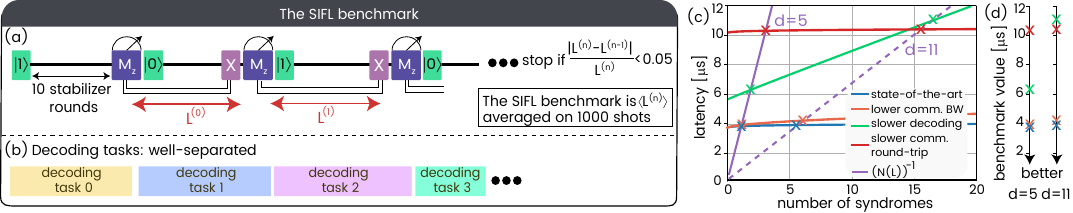}
    \caption{The Stead-state Inter-circuit Feed-forward Latency (SIFL) benchmark. (a) Logical circuit for measuring the SIFL benchmark. Double lines indicate conditional gates dependent on decoding; red arrow shows the FFL for the first two decoding tasks. The number of circuits is set by a stopping condition; SIFL is defined as the FFL of the final circuit, averaged over 1000 repetitions. (b) Decoding task view: tasks are sequential and isolated. The first task (yellow) runs a fixed 10-round stabilizer circuit; subsequent rounds to be decoded depend on their prior FFL (\(L^{(n-1)}\)). (c) Simulated FFL vs. number of decoded syndromes for various CDUs: state-of-the-art CDU (blue), and when modifying to lower communication bandwidth (orange), slower decoding (green), or slower communication round-trip (red). Parameters detailed in Table S2; state-of-the-art decoding latency is extracted from experimental data in \cite{liyanage2023scalable}, and the communication round trip and bandwidth from the DGX-Quantum parameters \cite{mohseni2024build}. Purple curves represent \((N(L))^{-1}\), derived from the error rates and circuit size for the SIFL 49 (d=5) and the SIFL 241 (d=11). The intersection between these curves and the FFL traces correspond to the steady-state operating points of the system. (d) SIFL results across systems and code distances, showing the impact of CDU characteristics.}
    \label{fig5}
\end{figure*}

The goal of the first benchmark is to quantify the minimal closed-loop latency of the CDU while ensuring tight integration, with all components performing their roles in synchrony. This benchmark includes two variants to evaluate first the minimal latency and second the additional latency due to the decoder's throughput or communication bandwidth for near-term circuits, but does not assess long-term scalability. Instead, it verifies that the CDU can perform all essential classical operations required for near-term non-Clifford QEC circuits. 

Using the mathematical notations presented above, the first variant of this benchmark (related to a distance-5 surface) is determined primarily by \(\tau_0\), which represents the minimal achievable latency dictated by the individual components. Contributors to \(\tau_0\) include the latencies of the controller’s analog-to-digital conversion, the communication channel, decoder initialization, and the fixed latency of the decoding algorithm itself. The second variant (related to a distance-11 surface) includes the evaluation of the decoding complexity or communication bandwidth, which relate to the parameter \(\alpha\).  

We call this benchmark the Steady-State Inter-circuit Feed-forward Latency (SIFL) benchmark which is based on a preliminary use-case in which the quantum resources are limited to a single logical qubit. Although a single surface is not sufficient for performing quantum computation, this benchmark stands as a first step towards the scalability of QEC quantum computation, suitable for checking the CDU capabilities with available decoding systems. 

The logical circuit implemented in the benchmark is shown in Fig. 5a. The controller initializes the surface (data qubits) in the \(|0\rangle\) state, performs 10 stabilizer rounds, and then measures the data qubits. The benchmark clock starts once the last sample of the last data qubit measurement signal is sampled by the controller (cf. the measurement timestamp in \cite{ella2023quantum}). The controller then immediately initializes the surface for the next circuit followed by execution of repeated stabilizer rounds until the decoding of the first circuit is done. Then, the controller applies the appropriate conditional feedback based on the decoded frame and measures the surface again. We measure latency of the first decoding task, denoted by \(L^{(0)}\), as the time from the start of the benchmark clock until the first sample of the conditional pulse is played from the controller (cf. the conditional pulse timestamp in \cite{ella2023quantum}).

To further verify the capability of the controller and the decoder to run in parallel, and to quantify the CDU's closed-loop, we do not stop the benchmark once the first FFL is measured (related to the decoding of 10 stabilizer rounds). We set the end of the benchmark to be the steady-state FFL. We define steady-state stopping condition as \(\frac{L^{(n)} - L^{(n-1)}}{L^{(n)}} < 0.05\), where \(L^{(n)}\) relates to an unknown decoding task \(n\), which contains an unknown number of stabilizer rounds. To avoid irregularities, we define the benchmark as the average of 1000 shots for which \(L^{(n)}\) is found. 

Lastly, we define the size of the surface for the benchmark to be surface 49 (distance-5) or surface 241 (distance-11), and refer to these benchmarks as Steady-state Inter-circuit FFL (SIFL) 49 or 241. The SIFL 49 benchmark includes minimal decoding computation, making it more biased towards the value of \(\tau_0\)\, which sums the controller, decoder, and two-way communication minimal latency. The SIFL 241 includes additional evaluation of the decoder's throughput complexity and the controller-decoder communication channel bandwidth.

Figs. 5c-d illustrate concrete examples of the SIFL benchmark values (5d) and their extraction method (5c). Panel (c) presents simulated FFL as a function of the number of syndromes to decode, comparing a state-of-the-art CDU (blue) to variations with reduced communication bandwidth (orange), slower decoding (green), and increased communication round-trip latency (red). The state-of-the-art decoding latency was extracted from the experimental data in Ref. \cite{liyanage2023scalable}, which demonstrated sub-linear scaling with increasing syndrome count. This decoding latency was added to the DGX-Quantum controller-decoder round-trip latency (3.8 \(\mu s\)) \cite{mohseni2024build}, and the multiplication of its bandwidth (64 \(Gb/s\)) with the number of stabilizer measurements to obtain the overall FFL of the CDU. The suboptimal parameters are presented in Table S2.

We note that these latencies are plotted as a function of the number of syndromes to be decoded. This choice for the x-axis is motivated by the fact that the latency of matching-based decoding algorithms depends on the number of matching events. The number of syndromes to be decoded is calculated through the number of stabilizer measurements (derived by the code distance and the number QEC rounds) times the probability that a measurement is a syndrome (0.025 for error probability of \(0.5\%\) \cite{barber2025real}). 

The benchmark values, indicated by crosses in Figs. 5c and 5d, are extracted from the intersection between the FFL curves and the inverse of the syndrome rate per latency, \((N(L))^{-1}\) (plotted in purple). This curve is derived from the benchmark definition, which is based on the QEC cycle time, code distance, and physical error rate. As in Fig. 4, the intersection point determines the steady-state latency and hence the corresponding benchmark value. At \(d=5\) these intersections show that the SIFL 49 benchmark is primarily determined by the minimal latency contributions of the classical components, including integration overheads.

\lstset{
    captionpos=b, 
    aboveskip=0pt, 
    belowskip=0pt
}

\begin{lstlisting}[caption={\textbf{Pseudocode for measuring the Stead-state Intercircuit Feed-forward Latency}. Green: elements of the configuration file. Blue: real-time classical variables. Red: pulse-level commands. Orange: pre-defined constants and functions. Bright Purple: control flow statement. Cyan: controller-decoder communication statement.Mustard: decoding algorithm. We note that the macros “stabilizer round”, “initialize surface”, and “measure surface” are defined in Listing S1.}]
rounds=10
error_probability=0.005 
round_time=1E-6 #microsecond
latency_limit=1 #second
averaging_loop=1000
decoding_algorithm=Union_find
send_to_decoder(algorithm=decoding_algorithm)
SIFL_of_sample=[]
initialize_simulation(error_probability,round_time)  
for i in range(averaging_loop):
 task_indx=0
 task_latency=[latency_limit]
 initialize_surface(q0, state=1)
 for j in range(rounds):
  ancilla_bits=stabilizer_round()
  send_to_decoder(ancilla_bits, task=task_indx)
 data_bits=measure_surface(q0, timestamp->tic)
 send_to_decoder(data_bits, task=task_indx)
 while task_latency[-1]<latency_limit:
  task_indx+=1
  initialize_surface(q0, state=1+ task_indx mod 2)
  [logical_result, decoding_recieved]=get_decoding_result(task=task_indx-1)
  while not decoding_recieved:
   ancilla_bits=stabilizer_round()
   send_to_decoder(ancilla_bits, task=task_indx)
   [logical_result, decoding_received]=get_decoding_result(task= task_indx-1)
  play_x(q0, timestamp->toc)
  task_latency.append(toc-tic)
  if abs(task_latency[-1]-task_latency[-2])<0.05*task_latency[-1]:
   SIFL_of_sample.append(task_latency[-1])
   break  
  data_bits=measure_surface(q0, error_probability, timestamp->tic) 
SIFL=mean(SIFL_of_sample)
\end{lstlisting}

At \(d=11\), the intersections reveal how the SIFL 241 benchmark begins to reflect the CDU’s throughput limits. Specifically, the additional time required for decoding (green curve) or data transfer (orange curve) increases the benchmark value compared to the SIFL 49. Notably, the benchmark of a CDU comprised by a linear decoder (green curve) and state-of-the-art controller and communication channel becomes worse than that of a CDU with high controller-decoder round trip value and sub-linear decoding (red curve). In a broader perspective, the benchmark extraction in Figs. 5c illustrates how the dynamic system diagram enables clear and comprehensive benchmarking of the entire classical stack involved in QEC computation.

The concrete definition of the SIFL benchmark is shown in Listing 1 as pseudocode. We write the variables and building block commands required to run a single-surface experiment and measure the SIFL benchmarks. The pulse-level statements and macros \textcolor{red}{\textbf{initialize surface}}, \textcolor{red}{\textbf{play}}, \textcolor{red}{\textbf{stabilizer round}}, and \textcolor{red}{\textbf{measure surface}} include a set of predefined physical RF pulses and measurements that are detailed in Listing S1 and are compatible with the representative use case discussed above. To create the separation between the classical hardware that we want to benchmark and the quantum hardware (that is, to be able to benchmark the classical hardware without quantum hardware), we add as an input to these operations the predefined parameters \textcolor{orange}{\textbf{error probability}} and \textcolor{orange}{\textbf{round time}} for the stabilizer round that can be controlled classically in the analogue input to the controller. 

Apart from the analog signals, we write the controller-decoder communication commands (\textcolor{cyan}{Cyan}) and decoding algorithm (\textcolor{mustard}{mustard}) explicitly. In addition, control-flow commands are essential since the controller is required to execute stabilizer rounds for an unknown number of rounds during the decoding. This operation is described by the \textcolor{purple}{\textbf{while}} loop where classical data is acquired during the decoding process. Interestingly, although we aim for a surface code, the pseudocode in Listing 1 can be used to benchmark any QEC stabilizer code. 

\subsection{MEDIUM-TERM BENCHMARK: STEADY-STATE SURGERY FEED-FORWARD LATENCY (SSFL)}

\begin{figure}[b]
    \centering
    \includegraphics[width=\linewidth]{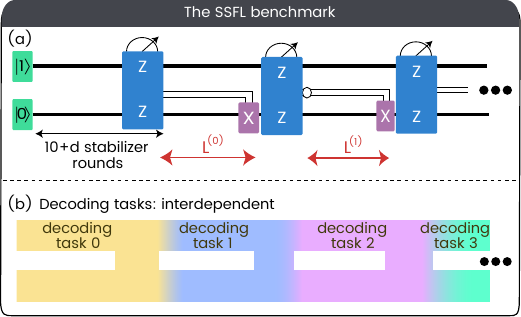}
    \caption{The Stead-state Surgery Feed-forward Latency (SSFL) benchmark. (a) Logical circuit for measuring the SSFL benchmark. The represents a conditional gate, and the red arrows represent the FFL of the first two decoding tasks where each decoding task ends with the termination of the lattice surgery. The number of decoding tasks is determined by the stopping condition. The SSFL benchmark is the feed-forward latency of the last decoding tasks, averaged by 1000 shots. The stabilizer rounds are not shown since they represent a logical identity. (b) The decoding task view of the SSFL benchmark, where each decoding task is not strictly separated from the other decoding tasks.}
    \label{fig6}
\end{figure}

To provide a benchmark that confronts the most demanding part for running non-Clifford circuits in terms of real-time decoding, we define a second benchmark. In this benchmark, we modify the SIFL benchmark so that the logical measurement will be a lattice surgery rather than a single surface measurement, as shown in Fig. 6 and defined in Listing 2.

The lattice surgery feed-forward is one of the main building blocks for performing non-Clifford gates (also in color codes \cite{landahl2014quantum}), and moreover, is the only type of feed-forward required for the entire quantum circuit. As in the previous benchmark, we evaluate the decoding throughput, the classical-quantum parallelization, and the speed of the overall integration, by defining the benchmark as a steady-state surgery FFL (SSFL). Each latency is defined from the moment one lattice surgery ends (where each surgery has d stabilizer rounds) until the moment the corresponding feed-forward is applied, while the stopping condition and averaging is similar to the SIFL benchmark. In this case, we define the benchmark for a distance-3 or distance-5 configuration with a single data qubit column (as in Fig.1b), containing 41 or 111 physical qubits.

\begin{lstlisting}[label={lst:ssfl}, caption={\textbf{Pseudocode for measuring the Stead-state Surgery Feed-forward Latency (SSFL)}. Color coding is the same as in Listing 1, the macros ``initialize surgery'' and ``terminate surgery'' are defined in Listing S1.}]
rounds=10
surgery_rounds=d
error_probability=0.005 
round_time=1E-6 #microsecond
latency_limit=1 #second
averaging_loop=1000
decoding_algorithm=Union_find
send_to_decoder(algorithm=decoding_algorithm)
initialize_simulation(error_probability, round_time)  
SSFL_of_sample=[]
for i in range(averaging_loop):
 task_indx=0
 task_latency=[latency_limit]
 initialize_surface(q0, state=1)
 initialize_surface(q1, state=0)
 for i in range(rounds):
  ancilla_bits=stabilizer_round()
   send_to_decoder(ancilla_bits, task=task_indx)
 while task_latency[-1]< latency_limit:
  initialize_surgery(q0, q1)
  for i in range(surgery_rounds):
    ancilla_bits=stabilizer_round()
    send_to_decoder(ancilla_bits, task=task_indx)
  surgery_data_bits=terminate_surgery((q0, q1), timestamp->tic)
  send_to_decoder(surgery_data_bits, task=task_indx)
  task_indx=task_indx+1
  [logical_result, decoding_received]=get_decoding_result(task=task_indx-1)
  while not decoding_recieved:
   ancilla_bits=stabilizer_round()
   send_to_decoder(ancilla_bits, task=task_indx)
   [logical_result, decoding_received]=get_decoding_result(task=task_indx-1)
  play_x(q1, timestamp->toc)
  task_latency.append(toc-tic)
  if abs(task_latency[-1]-task_latency[-2])<0.05*task_latency[-1]:
   SSFL_of_sample.append(task_latency[-1])
   break
SSFL=mean(SSFL_of_sample)
\end{lstlisting}

As in the SILF benchmark, the SSFL benchmark also includes idling stabilizer rounds until the feed-forward is applied (as in the circuit in Fig. 3 and Listing 2), so that the latency of the first decoding task will determine the amount of processing data of the second decoding task. However, a significant difference between the two benchmarks comes from the fact that the decoding tasks are not well-separated, as we show in Figures 6b. The second decoding task will depend on the syndromes from the first decoding task, or other boundary conditions that are transferred between decoding tasks to overcome a set of errors with syndromes in both tasks (illustrated by the connection between the tasks). As a result, we expect an additional demand on the CDU when implementing the SSFL benchmark compared to the SIFL benchmark, which will result in a latency expansion and an overall worse benchmark value. 

 Importantly, the biggest challenge in implementing this benchmark is that, to the best of our knowledge, it is currently impossible with the available decoding tools (such as PyMatching \cite{higgott2022pymatching} or Fusion-Blossom \cite{wu2023fusion}) to have a dependency between decoding tasks which share correlated syndromes. Hence, executing the SSFL benchmark will necessitate algorithmic development alongside control hardware improvements. 

\subsection{Generalization and Practical Use of Benchmarks}

To demonstrate the use of the proposed benchmarks, we chose a representative case that pushes classical hardware to its limits: superconducting qubit systems, which currently operate with the fastest clock cycles. As such, a CDU which reaches benchmark values in the \(\mu s\)-scale will be applicable to slower platforms such as trapped ions and Rydberg atom arrays. 

For consistency in evaluation, we define a specific decoding algorithm based on the widely used minimum-weight-perfect-matching (MWPM) as a representative example. Choosing a predefined decoding algorithm ensures that different CDUs perform comparable classical computations, producing consistent outputs for similar inputs. However, the benchmark definitions (listings 1 and 2) are parametrized, so they can be simply modified to compare different QEC code and decoding algorithms within the same hardware. In addition, the homogeneous physical error rate of \(0.5\%\) which we define can be modified to compare between different error rates and error models. 

The SIFL benchmark simplifies the requirements compared to those of non-Clifford gate execution, due to the separation between decoding tasks, as shown in Fig. 5b. This separation allows the SIFL benchmark to be implemented using various decoding algorithms that rely on well-defined task boundaries. As a result, the SIFL benchmark can already be used to evaluate current available hardware. 

The SSFL benchmark, however, requires real-time decoding of lattice surgery operations, a capability that has not yet been demonstrated. Nonetheless, we emphasize the importance of this benchmark, as it directly evaluates the specific functionality required from a CDU during the execution of non-Clifford QEC circuits. We anticipate that systems capable of supporting and validating this benchmark will become available in the coming years.

\section{DISCUSSION}
In this manuscript, we have explained the central roles of the CDU, the necessity for a low FFL, and suggested two near-term benchmarks for evaluating the CDU performance. The suggested benchmarks provide a first reduction of all CDU components into few holistic evaluation numbers. Although these benchmarks do not perform any useful quantum calculations, they are defined using representative configurations according to current state-of-the-art parameters. Moreover, they verify the ability of the CDU to support future state-of-the-art QEC experiments, setting the stage for scaling quantum hardware. 

These benchmarks, the commands within the pseudocodes, and the operation regime analysis, are general for most other QEC stabilizer codes which may have a different detailed physical pulse sequence and decoding algorithms. Eventually, these benchmarks are intended to enable real-time decoding with the highest performance in novel near-term QEC experiments with available quantum hardware.

In the long term, we expect that the benchmarks should be extended to incorporate the CDU's scalability, which will rely on the dominant role in parallelizing classical calculations within the CDU. The CDU will need to parallelize many decoding tasks to keep track of the frame of each logical qubit, and then merge decoding tasks during multi-surface lattice surgeries \cite{kurman2024controller}. In addition, it would be beneficial to consider indirect benchmarks for evaluating a QEC operation runtime, including the time it takes to run multiple shots, to compile logical circuits into T-gate operations \cite{heyfron2018efficient}, and to load the circuit parameters (e.g., to synthesize circuit waveform sequences). In this context, an interesting benchmark will be the time it takes to perform embedded calibrations for the circuit parameters optimization (e.g.,\cite{sivak2023real, barrett2023learning, klimov2024optimizing}). The quality of the optimization will eventually determine the physical error probability of the circuit which in turn gives rise to another possible set of benchmarks related to the decoder adaptation time. The decoder will be required to adapt its weights and estimates of the circuit noise (for example by using optimal noise estimation \cite{wagner2021optimal}) and to adapt to catastrophic events \cite{mcewen2022resolving}.

We foresee that optimizing for the benchmarks that we defined here while keeping high decoding accuracy will be at the core of quantum computation utility. Once the surgery latency is minimized, the effort will move towards minimizing the multi-surface surgery latency, and eventually the magic state distillation time. The distillation time is expected to determine de facto the whole quantum computation time for surface-code-based computation, which is currently the main expected route for reducing dramatically logical error rates. Finally, even if other promising QEC codes such as QLDPC codes\cite{pattison2023hierarchical}  or other unique codes \cite{ruiz2025ldpc} will become dominant, all properties and blueprints for CDU that we describe here will still stand. Thus, having clear benchmarks that combine the complete requirements of a CDU is a vital and fundamental component in the development of quantum computers. 

\section*{Acknowledgment}

This work was supported in part Horizon Europe programme HORIZON-CL4-2021-DIGITAL-EMERGING-01-30 via the project 101070144 (EuRyQa)

\bibliographystyle{IEEEtran} 
\bibliography{tqe}

\begin{thebibliography}{10}
\providecommand{\url}[1]{#1}
\csname url@samestyle\endcsname
\providecommand{\newblock}{\relax}
\providecommand{\bibinfo}[2]{#2}
\providecommand{\BIBentrySTDinterwordspacing}{\spaceskip=0pt\relax}
\providecommand{\BIBentryALTinterwordstretchfactor}{4}
\providecommand{\BIBentryALTinterwordspacing}{\spaceskip=\fontdimen2\font plus
\BIBentryALTinterwordstretchfactor\fontdimen3\font minus \fontdimen4\font\relax}
\providecommand{\BIBforeignlanguage}[2]{{%
\expandafter\ifx\csname l@#1\endcsname\relax
\typeout{** WARNING: IEEEtran.bst: No hyphenation pattern has been}%
\typeout{** loaded for the language `#1'. Using the pattern for}%
\typeout{** the default language instead.}%
\else
\language=\csname l@#1\endcsname
\fi
#2}}
\providecommand{\BIBdecl}{\relax}
\BIBdecl

\bibitem{shor1996fault}
P.~W. Shor, ``Fault-tolerant quantum computation,'' in \emph{Proceedings of 37th conference on foundations of computer science}.\hskip 1em plus 0.5em minus 0.4em\relax IEEE, 1996, pp. 56--65.

\bibitem{knill1998resilient}
E.~Knill, R.~Laflamme, and W.~H. Zurek, ``Resilient quantum computation: error models and thresholds,'' \emph{Proceedings of the Royal Society of London. Series A: Mathematical, Physical and Engineering Sciences}, vol. 454, no. 1969, pp. 365--384, 1998.

\bibitem{feynman2018simulating}
R.~P. Feynman, ``Simulating physics with computers,'' in \emph{Feynman and computation}.\hskip 1em plus 0.5em minus 0.4em\relax cRc Press, 2018, pp. 133--153.

\bibitem{shor1999polynomial}
P.~W. Shor, ``Polynomial-time algorithms for prime factorization and discrete logarithms on a quantum computer,'' \emph{SIAM review}, vol.~41, no.~2, pp. 303--332, 1999.

\bibitem{gottesman1997stabilizer}
D.~Gottesman, \emph{Stabilizer codes and quantum error correction}.\hskip 1em plus 0.5em minus 0.4em\relax California Institute of Technology, 1997.

\bibitem{preskill1998reliable}
J.~Preskill, ``Reliable quantum computers,'' \emph{Proceedings of the Royal Society of London. Series A: Mathematical, Physical and Engineering Sciences}, vol. 454, no. 1969, pp. 385--410, 1998.

\bibitem{campbell2017roads}
E.~T. Campbell, B.~M. Terhal, and C.~Vuillot, ``Roads towards fault-tolerant universal quantum computation,'' \emph{Nature}, vol. 549, no. 7671, pp. 172--179, 2017.

\bibitem{demarti2024decoding}
A.~deMarti iOlius, P.~Fuentes, R.~Or{\'u}s, P.~M. Crespo, and J.~E. Martinez, ``Decoding algorithms for surface codes,'' \emph{Quantum}, vol.~8, p. 1498, 2024.

\bibitem{ofek2016extending}
N.~Ofek, A.~Petrenko, R.~Heeres, P.~Reinhold, Z.~Leghtas, B.~Vlastakis, Y.~Liu, L.~Frunzio, S.~M. Girvin, L.~Jiang \emph{et~al.}, ``Extending the lifetime of a quantum bit with error correction in superconducting circuits,'' \emph{Nature}, vol. 536, no. 7617, pp. 441--445, 2016.

\bibitem{devulapalli2024quantum}
D.~Devulapalli, E.~Schoute, A.~Bapat, A.~M. Childs, and A.~V. Gorshkov, ``Quantum routing with teleportation,'' \emph{Physical Review Research}, vol.~6, no.~3, p. 033313, 2024.

\bibitem{baumer2024efficient}
E.~B{\"a}umer, V.~Tripathi, D.~S. Wang, P.~Rall, E.~H. Chen, S.~Majumder, A.~Seif, and Z.~K. Minev, ``Efficient long-range entanglement using dynamic circuits,'' \emph{PRX Quantum}, vol.~5, no.~3, p. 030339, 2024.

\bibitem{foss2023experimental}
M.~Foss-Feig, A.~Tikku, T.-C. Lu, K.~Mayer, M.~Iqbal, T.~M. Gatterman, J.~A. Gerber, K.~Gilmore, D.~Gresh, A.~Hankin \emph{et~al.}, ``Experimental demonstration of the advantage of adaptive quantum circuits,'' \emph{arXiv preprint arXiv:2302.03029}, 2023.

\bibitem{ella2023quantum}
L.~Ella, L.~Leandro, O.~Wertheim, Y.~Romach, L.~Schlipf, R.~Szmuk, Y.~Knol, N.~Ofek, I.~Sivan, and Y.~Cohen, ``Quantum-classical processing and benchmarking at the pulse-level,'' \emph{arXiv preprint arXiv:2303.03816}, 2023.

\bibitem{ai2024quantum}
G.~Q. AI \emph{et~al.}, ``Quantum error correction below the surface code threshold,'' \emph{Nature}, vol. 638, no. 8052, p. 920, 2024.

\bibitem{bluvstein2024logical}
D.~Bluvstein, S.~J. Evered, A.~A. Geim, S.~H. Li, H.~Zhou, T.~Manovitz, S.~Ebadi, M.~Cain, M.~Kalinowski, D.~Hangleiter \emph{et~al.}, ``Logical quantum processor based on reconfigurable atom arrays,'' \emph{Nature}, vol. 626, no. 7997, pp. 58--65, 2024.

\bibitem{gottesman1998heisenberg}
D.~Gottesman, ``The heisenberg representation of quantum computers,'' \emph{arXiv preprint quant-ph/9807006}, 1998.

\bibitem{dawson2005solovay}
C.~M. Dawson and M.~A. Nielsen, ``The solovay-kitaev algorithm,'' \emph{arXiv preprint quant-ph/0505030}, 2005.

\bibitem{riesebos2017pauli}
L.~Riesebos, X.~Fu, S.~Varsamopoulos, C.~G. Almudever, and K.~Bertels, ``Pauli frames for quantum computer architectures,'' in \emph{Proceedings of the 54th Annual Design Automation Conference 2017}, 2017, pp. 1--6.

\bibitem{terhal2015quantum}
B.~M. Terhal, ``Quantum error correction for quantum memories,'' \emph{Reviews of Modern Physics}, vol.~87, no.~2, pp. 307--346, 2015.

\bibitem{caune2024demonstrating}
L.~Caune, L.~Skoric, N.~S. Blunt, A.~Ruban, J.~McDaniel, J.~A. Valery, A.~D. Patterson, A.~V. Gramolin, J.~Majaniemi, K.~M. Barnes \emph{et~al.}, ``Demonstrating real-time and low-latency quantum error correction with superconducting qubits,'' \emph{arXiv preprint arXiv:2410.05202}, 2024.

\bibitem{liyanage2023scalable}
N.~Liyanage, Y.~Wu, A.~Deters, and L.~Zhong, ``Scalable quantum error correction for surface codes using fpga,'' in \emph{2023 IEEE International Conference on Quantum Computing and Engineering (QCE)}, vol.~1.\hskip 1em plus 0.5em minus 0.4em\relax IEEE, 2023, pp. 916--927.

\bibitem{das2022afs}
P.~Das, C.~A. Pattison, S.~Manne, D.~M. Carmean, K.~M. Svore, M.~Qureshi, and N.~Delfosse, ``Afs: Accurate, fast, and scalable error-decoding for fault-tolerant quantum computers,'' in \emph{2022 IEEE International Symposium on High-Performance Computer Architecture (HPCA)}.\hskip 1em plus 0.5em minus 0.4em\relax IEEE, 2022, pp. 259--273.

\bibitem{maurya2024managing}
S.~Maurya and S.~Tannu, ``Managing classical processing requirements for quantum error correction,'' \emph{arXiv preprint arXiv:2406.17995}, 2024.

\bibitem{wack2110quality}
A.~Wack, H.~Paik, A.~Javadi-Abhari, P.~Jurcevic, I.~Faro, J.~M. Gambetta, and B.~R. Johnson, ``Quality, speed, and scale: three key attributes to measure the performance of near-term quantum computers. 2021. doi: 10.48550,'' \emph{arXiv preprint arXiv.2110.14108}.

\bibitem{battistel2023real}
F.~Battistel, C.~Chamberland, K.~Johar, R.~W. Overwater, F.~Sebastiano, L.~Skoric, Y.~Ueno, and M.~Usman, ``Real-time decoding for fault-tolerant quantum computing: Progress, challenges and outlook,'' \emph{Nano Futures}, vol.~7, no.~3, p. 032003, 2023.

\bibitem{wu2023fusion}
Y.~Wu and L.~Zhong, ``Fusion blossom: Fast mwpm decoders for qec,'' in \emph{2023 IEEE International Conference on Quantum Computing and Engineering (QCE)}, vol.~1.\hskip 1em plus 0.5em minus 0.4em\relax IEEE, 2023, pp. 928--938.

\bibitem{barber2025real}
B.~Barber, K.~M. Barnes, T.~Bialas, O.~Bu{\u{g}}dayc{\i}, E.~T. Campbell, N.~I. Gillespie, K.~Johar, R.~Rajan, A.~W. Richardson, L.~Skoric \emph{et~al.}, ``A real-time, scalable, fast and resource-efficient decoder for a quantum computer,'' \emph{Nature Electronics}, pp. 1--8, 2025.

\bibitem{dennis2002topological}
E.~Dennis, A.~Kitaev, A.~Landahl, and J.~Preskill, ``Topological quantum memory,'' \emph{Journal of Mathematical Physics}, vol.~43, no.~9, pp. 4452--4505, 2002.

\bibitem{fowler2012surface}
A.~G. Fowler, M.~Mariantoni, J.~M. Martinis, and A.~N. Cleland, ``Surface codes: Towards practical large-scale quantum computation,'' \emph{Physical Review A—Atomic, Molecular, and Optical Physics}, vol.~86, no.~3, p. 032324, 2012.

\bibitem{horsman2012surface}
D.~Horsman, A.~G. Fowler, S.~Devitt, and R.~Van~Meter, ``Surface code quantum computing by lattice surgery,'' \emph{New Journal of Physics}, vol.~14, no.~12, p. 123011, 2012.

\bibitem{chamberland2022universal}
C.~Chamberland and E.~T. Campbell, ``Universal quantum computing with twist-free and temporally encoded lattice surgery,'' \emph{PRX Quantum}, vol.~3, no.~1, p. 010331, 2022.

\bibitem{litinski2019game}
D.~Litinski, ``A game of surface codes: Large-scale quantum computing with lattice surgery,'' \emph{Quantum}, vol.~3, p. 128, 2019.

\bibitem{higgott2023improved}
O.~Higgott, T.~C. Bohdanowicz, A.~Kubica, S.~T. Flammia, and E.~T. Campbell, ``Improved decoding of circuit noise and fragile boundaries of tailored surface codes,'' \emph{Physical Review X}, vol.~13, no.~3, p. 031007, 2023.

\bibitem{li2015magic}
Y.~Li, ``A magic state’s fidelity can be superior to the operations that created it,'' \emph{New Journal of Physics}, vol.~17, no.~2, p. 023037, 2015.

\bibitem{gidney2023cleaner}
C.~Gidney, ``Cleaner magic states with hook injection,'' \emph{arXiv preprint arXiv:2302.12292}, 2023.

\bibitem{kolmogorov2009blossom}
V.~Kolmogorov, ``Blossom v: a new implementation of a minimum cost perfect matching algorithm,'' \emph{Mathematical Programming Computation}, vol.~1, pp. 43--67, 2009.

\bibitem{delfosse2021almost}
N.~Delfosse and N.~H. Nickerson, ``Almost-linear time decoding algorithm for topological codes,'' \emph{Quantum}, vol.~5, p. 595, 2021.

\bibitem{chubb2021general}
C.~T. Chubb, ``General tensor network decoding of 2d pauli codes,'' \emph{arXiv preprint arXiv:2101.04125}, 2021.

\bibitem{das2022lilliput}
P.~Das, A.~Locharla, and C.~Jones, ``Lilliput: a lightweight low-latency lookup-table decoder for near-term quantum error correction,'' in \emph{Proceedings of the 27th ACM International Conference on Architectural Support for Programming Languages and Operating Systems}, 2022, pp. 541--553.

\bibitem{meinerz2022scalable}
K.~Meinerz, C.-Y. Park, and S.~Trebst, ``Scalable neural decoder for topological surface codes,'' \emph{Physical Review Letters}, vol. 128, no.~8, p. 080505, 2022.

\bibitem{higgott2022pymatching}
O.~Higgott, ``Pymatching: A python package for decoding quantum codes with minimum-weight perfect matching,'' \emph{ACM Transactions on Quantum Computing}, vol.~3, no.~3, pp. 1--16, 2022.

\bibitem{gidney2021stim}
C.~Gidney, ``Stim: a fast stabilizer circuit simulator,'' \emph{Quantum}, vol.~5, p. 497, 2021.

\bibitem{haah2018codes}
J.~Haah and M.~B. Hastings, ``Codes and protocols for distilling $ t $, controlled-$ s $, and toffoli gates,'' \emph{Quantum}, vol.~2, p.~71, 2018.

\bibitem{litinski2019magic}
D.~Litinski, ``Magic state distillation: Not as costly as you think,'' \emph{Quantum}, vol.~3, p. 205, 2019.

\bibitem{fu2019control}
X.~Fu, L.~Lao, K.~Bertels, and C.~G. Almudever, ``A control microarchitecture for fault-tolerant quantum computing,'' \emph{Microprocessors and Microsystems}, vol.~70, pp. 21--30, 2019.

\bibitem{kurman2024controller}
Y.~Kurman, L.~Ella, N.~Halay, O.~Wertheim, and Y.~Cohen, ``Controller-decoder system requirements derived by implementing shor's algorithm with surface code,'' \emph{arXiv preprint arXiv:2412.00289}, 2024.

\bibitem{akahoshi2024partially}
Y.~Akahoshi, K.~Maruyama, H.~Oshima, S.~Sato, and K.~Fujii, ``Partially fault-tolerant quantum computing architecture with error-corrected clifford gates and space-time efficient analog rotations,'' \emph{PRX quantum}, vol.~5, no.~1, p. 010337, 2024.

\bibitem{alligood1998chaos}
K.~T. Alligood, T.~D. Sauer, J.~A. Yorke, and D.~Chillingworth, ``Chaos: an introduction to dynamical systems,'' \emph{SIAM Review}, vol.~40, no.~3, pp. 732--732, 1998.

\bibitem{higgott2025sparse}
O.~Higgott and C.~Gidney, ``Sparse blossom: correcting a million errors per core second with minimum-weight matching,'' \emph{Quantum}, vol.~9, p. 1600, 2025.

\bibitem{mohseni2024build}
M.~Mohseni, A.~Scherer, K.~G. Johnson, O.~Wertheim, M.~Otten, N.~A. Aadit, K.~M. Bresniker, K.~Y. Camsari, B.~Chapman, S.~Chatterjee \emph{et~al.}, ``How to build a quantum supercomputer: Scaling challenges and opportunities,'' \emph{arXiv preprint arXiv:2411.10406}, 2024.

\bibitem{landahl2014quantum}
A.~J. Landahl and C.~Ryan-Anderson, ``Quantum computing by color-code lattice surgery,'' \emph{arXiv preprint arXiv:1407.5103}, 2014.

\bibitem{heyfron2018efficient}
L.~E. Heyfron and E.~T. Campbell, ``An efficient quantum compiler that reduces t count,'' \emph{Quantum Science and Technology}, vol.~4, no.~1, p. 015004, 2018.

\bibitem{sivak2023real}
V.~V. Sivak, A.~Eickbusch, B.~Royer, S.~Singh, I.~Tsioutsios, S.~Ganjam, A.~Miano, B.~Brock, A.~Ding, L.~Frunzio \emph{et~al.}, ``Real-time quantum error correction beyond break-even,'' \emph{Nature}, vol. 616, no. 7955, pp. 50--55, 2023.

\bibitem{barrett2023learning}
C.~N. Barrett, A.~H. Karamlou, S.~E. Muschinske, I.~T. Rosen, J.~Braum{\"u}ller, R.~Das, D.~K. Kim, B.~M. Niedzielski, M.~Schuldt, K.~Serniak \emph{et~al.}, ``Learning-based calibration of flux crosstalk in transmon qubit arrays,'' \emph{Physical Review Applied}, vol.~20, no.~2, p. 024070, 2023.

\bibitem{klimov2024optimizing}
P.~V. Klimov, A.~Bengtsson, C.~Quintana, A.~Bourassa, S.~Hong, A.~Dunsworth, K.~J. Satzinger, W.~P. Livingston, V.~Sivak, M.~Y. Niu \emph{et~al.}, ``Optimizing quantum gates towards the scale of logical qubits,'' \emph{Nature Communications}, vol.~15, no.~1, p. 2442, 2024.

\bibitem{wagner2021optimal}
T.~Wagner, H.~Kampermann, D.~Bru{\ss}, and M.~Kliesch, ``Optimal noise estimation from syndrome statistics of quantum codes,'' \emph{Physical review research}, vol.~3, no.~1, p. 013292, 2021.

\bibitem{mcewen2022resolving}
M.~McEwen, L.~Faoro, K.~Arya, A.~Dunsworth, T.~Huang, S.~Kim, B.~Burkett, A.~Fowler, F.~Arute, J.~C. Bardin \emph{et~al.}, ``Resolving catastrophic error bursts from cosmic rays in large arrays of superconducting qubits,'' \emph{Nature Physics}, vol.~18, no.~1, pp. 107--111, 2022.

\bibitem{pattison2023hierarchical}
C.~A. Pattison, A.~Krishna, and J.~Preskill, ``Hierarchical memories: simulating quantum ldpc codes with local gates,'' \emph{arXiv preprint arXiv:2303.04798}, 2023.

\bibitem{ruiz2025ldpc}
D.~Ruiz, J.~Guillaud, A.~Leverrier, M.~Mirrahimi, and C.~Vuillot, ``Ldpc-cat codes for low-overhead quantum computing in 2d,'' \emph{Nature Communications}, vol.~16, no.~1, p. 1040, 2025.

\end{thebibliography}

\EOD
\end{document}